\documentclass[prd,aps,amsfonts,preprint,floats,tightenlines,floatfix,
nofootinbib,showpacs,showkeys]{revtex4}
\usepackage{exscale}                  % correct scaling of math-symbols
\usepackage[intlimits]{amsmath}       % improved mathematical typesetting
\usepackage{amsfonts}
\usepackage{amssymb,amscd}
\usepackage{epsfig}
\usepackage{wrapfig}

\newcommand{\si}{\sigma}
\newcommand{\ovs}{\overline{\sigma}}
\newcommand{\ovc}{\overline{c}}
\newcommand{\oG}{\tilde{\Gamma}}

\begin{document}

\date{\today}
\title{

%\hspace*{\fill}{\small\sf http://arXiv.org/abs/hep-ph/0411052}\\[4mm]

Infrared behavior of the ghost-gluon vertex in Landau gauge Yang--Mills theory}

\author{W.\ Schleifenbaum
\footnote{Present address: Institute of Theoretical Physics, U.\ of T\"ubingen, 
D-72070 T\"ubingen, Germany},}
\affiliation{Institute of Nuclear Physics, Technical University Darmstadt,
 D-64289 Darmstadt, Germany}

\author{A.\ Maas
\footnote{Present address: Gesellschaft f\"ur Schwerionenforschung, D-64291 Darmstadt, Germany},}
\affiliation{Institute of Nuclear Physics, Technical University Darmstadt,
 D-64289 Darmstadt, Germany}

\author{J.\ Wambach,}
\affiliation{Gesellschaft f\"ur Schwerionenforschung,
 D-64291 Darmstadt,
  Germany}
\affiliation{Institute of Nuclear Physics, Technical University Darmstadt,
 D-64289 Darmstadt, Germany}

\author{R.\ Alkofer
\footnote{Present address: Institute of Physics, University Graz,
Universit\"atsplatz 5, A-8010 Graz, Austria.}}
\affiliation{Institute of Theoretical Physics, U.\ of T\"ubingen, 
D-72070 T\"ubingen, Germany}

\begin{abstract}
%\normalsize
%\noindent
A semi-perturbative calculation of the ghost-gluon vertex in Landau gauge
Yang-Mills theory in four and three Euclidean space-time dimensions is 
presented. Non-perturbative gluon and ghost propagators are employed, 
which have previously been calculated from a truncated
set of Dyson--Schwinger equations and which are in qualitative and quantitative
agreement with corresponding lattice results. Our results for the
ghost-gluon vertex show only relatively small deviations from the tree-level
one in agreement with recent lattice data. In particular, we do
not see any sign for a singular behavior of the ghost-gluon vertex in the
infrared.
\end{abstract} 
%\vspace{1.5cm} 

\pacs{12.38.Aw, 14.70.Dj, 12.38.Lg, 11.15.Tk, 02.30.Rz, 11-10.Kk}
\keywords{Strong QCD, Ghost-Gluon
Vertex, Gluon Propagator, Dyson--Schwinger eqs., Infrared behavior}
\maketitle
%\noindent
%{\it Keywords:} Strong QCD, Ghost-Gluon
%Vertex, Gluon Propagator, Dyson--Schwinger eqs.,\\
%\phantom{\it Keywords:}  Infrared behavior

\section{Introduction}

The infrared behavior of QCD Green functions is of fundamental interest. 
In addition, these functions provide an important input for many
calculations in hadron physics, for recent reviews see {\it e.g.\/}
\cite{Maris:2003vk,Alkofer:2000wg,Roberts:2000aa}. As infrared singularities
are anticipated for some of these Green functions, non-perturbative continuum 
methods are needed to complement the knowledge gained in lattice Monte-Carlo
calculations. Studies using different techniques, such as 
Dyson-Schwinger equations (see {\it e.g.\/}  
\cite{Alkofer:2000wg,Fischer:2003rp,Alkofer:2003jj} and references therein),
renormalization group methods \cite{Pawlowski:2003hq}, stochastic quantization
\cite{Zwanziger:2003cf}, and lattice Monte-Carlo calculations (see {\it e.g.\/}
\cite{Sternbeck:2004xr,Cucchieri:2004mf,Oliveira:2004gy,Bowman:2004jm,Gattnar:2004bf}
and references therein)  have provided an unified picture of the infrared 
behavior of propagators in Landau gauge QCD in recent years. In this context the 
propagator of the Faddeev--Popov ghosts is of special interest: In the Landau gauge 
this propagator is infrared enhanced and diverges more strongly than $1/k^2$ for
$k^2\to 0$. On the one hand, this reflects the Zwanziger-Gribov horizon
condition \cite{Zwanziger:2003cf,Zwanziger:1993dh,Gribov:1977wm}. On the other
hand, in the Landau gauge it enforces the Kugo--Ojima confinement criterion
\cite{Kugo:1995km,Kugo:gm}. The accompanying infrared suppression of the gluon
propagator relates to positivity violation for transverse gluons by imposing a
cut in the gluon propagator \cite{Alkofer:2003jj}. This resolves an old puzzle
already encountered in perturbation theory, which has led to the 
Oehme-Zimmermann superconvergence relations \cite{Oehme:bj}. 

These Landau-gauge studies are complemented by similar ones in the Coulomb
gauge. Also in this gauge the infrared behavior of propagators is related to
the Gribov problem and confinement \cite{Feuchter:2004gb}. In addition, it has
been shown that center vortices play a crucial role in the infrared enhancement
of ghosts \cite{Greensite:2004ur}. Thus, the following picture emerges: Degrees of
freedom belonging to the indefinite-metric part of state space like the ghosts
in Landau gauge or ghosts and Coulomb gluons in Coulomb gauge are infrared
enhanced. This infrared enhancement is related directly to an effective cutoff
at the first Gribov horizon \cite{Zwanziger:2003cf,Gribov:1977wm}. The
corresponding ``excitations'' are confining in that they mediate long-range
correlations. Transverse gluons, on the other hand, are confined by these
modes. The infrared part of the transverse gluon propagator is strongly
suppressed. Besides an intuitive picture of confinement, this also provides a
formal line of reasoning: Violations of positivity remove these states from the
$S$ matrix.  Based on the relation of this picture to center vortices
\cite{Greensite:2004ur} it seems natural to speculate about the importance of
topological field configurations in this context.

Although the picture, emerging from different methods described above, is in
itself consistent and thus convincing, it is not yet complete. Finite-volume
effects prevent lattice calculations to explore the extreme infrared.
Functional continuum-based methods on the other hand necessarily involve
truncations and the related errors are hard to control. For the 
functional methods the Landau gauge is advantageous due to its
non-renormalization of the ghost-gluon vertex~\cite{Taylor:ff,Lerche:2002ep}.
To all orders in perturbation theory, the Landau gauge ghost-gluon vertex does
not develop a genuine ultraviolet divergence, and especially for vanishing incoming
ghost momentum it stays bare. Furthermore, it has been argued that, in the
extreme infrared, the gauge fixing term dominates over the Yang-Mills action
\cite{Zwanziger:2003cf}. Therefore, the infrared behavior of all Green
functions is expected to be dominated  by contributions involving ghosts. This
hypothesis has been tested for the gluon and ghost propagators and has
proven to be correct, thus alleviating very strongly the issue of truncation
induced errors. At this point, a truly non-perturbative investigation of the
ghost-gluon vertex has a twofold aim: First, it will add a further test of
ghost dominance in the infrared. Second, and more importantly, the result is
crucial to assess the validity of recent investigations based on functional
methods as all but the very first investigations\footnote{In these studies a
ghost-gluon vertex which is an approximate solution to the corresponding
Slavnov-Taylor  identity has been employed.} \cite{vonSmekal:1997is,Ellwanger:1996wy} 
used a bare ghost-gluon vertex. Thus, we will present a semi-perturbative calculation
of the ghost-gluon vertex based on its Dyson-Schwinger equation (DSE). For this
project, it has proven advantageous that lattice results for
the Landau gauge ghost-gluon vertex have been published very recently
\cite{Cucchieri:2004sq}. We will compare our predictions to these
data.

This paper is organized as follows: To make it reasonably self-contained we
will briefly discuss the non-perturbative gluon- and ghost propagators as they 
emerge from the solutions of their DSE's, truncated at the level
of propagators. Then, a truncation for the DSE of the 
ghost-gluon vertex will be given. We then discuss the results of
a semi-perturbative evaluation of this vertex. To this end, two types of input
for the vertex to be calculated are used. This, and the comparison to lattice
results, provides strong evidence that the full, non-perturbative ghost-gluon
vertex is very close to the tree-level one for all momenta.

\section{Gluon- and ghost propagators in Landau gauge QCD}

Yang-Mills theory in $D$-dimensional Euclidean spacetime in the Landau gauge is described by the Lagrangian \cite{Alkofer:2000wg}
\begin{eqnarray}
{\cal L}=&=&\frac{1}{4}F_{\mu\nu}^aF_{\mu\nu}^a+\bar c^a \partial_\mu D_\mu^{ab} c^b\label{lagrange}\\
F^a_{\mu\nu}&=&\partial_\mu A_\nu^a-\partial_\nu A_\mu^a-g_Df^{abc}A_\mu^bA_\nu^c\nonumber\\
D_\mu^{ab}&=&\delta^{ab}\partial_\mu+g_Df^{abc}A_\mu^c\nonumber~,
\end{eqnarray}
where $F_{\mu\nu}^a$ denotes the field strength tensor, $D_\mu^{ab}$ the covariant derivative, $g_D$ the $D$-dimensional gauge coupling, and $f^{abc}$ the structure constants of the gauge group. $A_\mu^a$ is the gluon field and $\bar c^a$ and $c^a$  are the Faddeev-Popov ghost fields, describing part of the intermediate states of the gluon field.

Within this framework, in Euclidean momentum space the Landau gauge gluon and ghost propagators, 
$ D_{\mu \nu}(p)$ and $D_G(p)$, can be generically written as
\begin{equation}
  D_{\mu \nu}(p,\mu^2) = \left(\delta_{\mu \nu} - \frac{p_\mu
      p_\nu}{p^2} \right) \frac{Z(p^2,\mu^2)}{p^2} \, ,
  \qquad
  D_G(p,\mu^2) = - \frac{G(p^2,\mu^2)}{p^2} \,,
  \label{g_prop}
\end{equation}
where $\mu^2$ denotes the renormalization scale, and $Z(p^2,\mu^2)$ and
$G(p^2,\mu^2)$ are the gluon and ghost dressing functions. They can be
determined from a solution of their DSE's
\cite{Fischer:2003rp,Fischer:2002hn,Maas:2004se} 
using a well-established truncation scheme 
\cite{vonSmekal:1997is,Fischer:2002hn}. 
A recent comparison of these solutions to the corresponding lattice results
can be found in ref.\ \cite{Fischer:2004ym,Maas:2004ec}. 
In the infrared, {\it i.e.\/} for infinitesimally small $p^2$, these equations
can be solved analytically 
\cite{Zwanziger:2003cf,Lerche:2002ep,vonSmekal:1997is} 
and one finds simple power laws,
\begin{equation}
  Z(p^2,\mu^2) \sim (p^2)^{2\kappa+2-D/2} \; , \qquad
  G(p^2,\mu^2) \sim (p^2)^{-\kappa}\; ,
  \label{g-power}
\end{equation}
for the gluon- and ghost dressing function with exponents related to
each other and to the dimensionality $D$. Here $\kappa$ is an irrational number,
$\kappa\approx 0.595$ for $D=4$ and $\kappa\approx 0.398$ for $D=3$
\cite{Lerche:2002ep,Zwanziger:2001kw}. 
These analytical results depend slightly on the
truncation scheme\footnote{For $D=4$  one can show, independent of any 
truncation, that $\kappa>0$ \cite{Watson:2001yv}.} 
\cite{Lerche:2002ep,Maas:2004se}. 
As mentioned above, this is in agreement with the Kugo-Ojima
confinement criterion and Zwanziger's horizon condition.

In four space-time dimensions 
the ghost- and gluon dressing functions can be used to define
a non-perturbative running coupling \cite{vonSmekal:1997is}
\begin{equation}
\alpha(p^2) = \alpha(\mu^2) \: G^2(p^2,\mu^2) \: Z(p^2,\mu^2)~.
  \label{alpha}
\end{equation}
Due to the ultraviolet finiteness of the ghost-gluon vertex in Landau gauge,
no vertex function appears in this definition.
Note that the r.h.s.\ of eq.\ (\ref{alpha}) is a renormalization group
invariant, and thus $\alpha(p^2)$ does not depend on the renormalization point. From the analytical power laws (\ref{g-power}) one infers
that the coupling has a fixed point in the infrared, given by $\alpha(0)\approx 8.92 / N_c$.
The infrared dominance of the ghosts imply 
that $\alpha(0)$ depends only weakly on the dressing of the ghost-gluon
vertex and not at all on other vertex functions \cite{Lerche:2002ep}.

In the following, for the calculation of the ghost-gluon vertex in four
space-time dimensions the pointwise accurate fit \cite{Fischer:2002hn}
\begin{eqnarray}
\alpha(p^2) = \frac{\alpha(0)}{\ln(e+a_1 p^{2a_2}+b_1p^{2b_2})}\; , \:\:\: &&
R(p^2) = \frac{c p^{2\kappa}+dp^{4\kappa}}{1+ c p^{2\kappa}+dp^{4\kappa}}\; , \nonumber\\
Z(p^2) = \left( \frac{\alpha(p^2)}{\alpha(\mu^2)} \right)^{1+2\delta} R^2(p^2)\;
, \:\:\: &&
G(p^2) = \left( \frac{\alpha(p^2)}{\alpha(\mu^2)} \right)^{-\delta} R^{-1}(p^2)
\; ,
\label{fit}
\end{eqnarray}
will be used. It employs
fitting parameters $a_1,a_2,b_1,b_2$ and $c,d$ for the running coupling
$\alpha(p^2)$ and the auxiliary function $R(p^2)$, respectively. Here, $\delta=-9/44$, 
is the anomalous dimension of the ghost dressing function and 
$\alpha(\mu^2=(1.31\;\mbox{GeV})^2)=0.9676$. The six parameters of the fit are given 
by $a_1=5.292\; \mbox{GeV}^{-2a_2}$, $a_2=2.324$,  
$b_1=0.034\; \mbox{GeV}^{-2b_2}$, $b_2=3.169$,
$c=1.8934\; \mbox{GeV}^{-2\kappa}$ and  $d=4.6944\; \mbox{GeV}^{-4\kappa}$. 
For the calculation in $D=3$, the numerical results 
for $G(p^2)$ and $Z(p^2)$ \cite{Maas:2004se} are directly used.

\section{Ghost-Gluon Vertex}

\begin{figure}[t]
\begin{center}
\epsfig{file=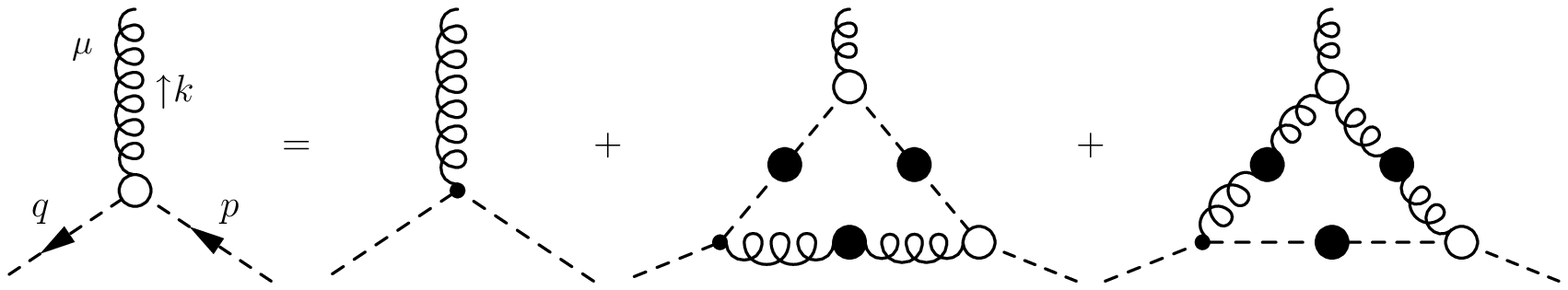,width=0.9\linewidth}
\end{center}
\caption{The truncated DSE for the ghost-gluon vertex. 
Dotted lines denote ghosts, and wiggly lines gluons. 
Lines with a dot indicate full
propagators. Vertices with small black dots represent bare vertices and open
circles represent full vertices. Contributions from the ghost-gluon 
scattering kernel have been neglected.} 
\label{figvertex}
\end{figure}

In the Landau gauge, the most general tensor structure of the ghost-gluon vertex 
with gluon momentum $k$ and ghost momenta $p$ and $q$ is given by
\begin{equation}
\Gamma^{abc}_\mu(k;q,p)=ig_D\left(q_\mu \left(f^{abc}+A^{abc}(k^2;q^2,p^2)\right)+
k_\mu B^{abc}(k^2;q^2,p^2)\right)\; ,
\label{tensor}
\end{equation}
where $A^{abc}$ and $B^{abc}$ are scalar functions describing the deviation 
from the tree-level form, and $g_D$ is the coupling constant. As there is no 
indication for a color structure different from the one occurring in perturbation 
theory~\cite{Alkofer:2000wg} we assume that 
$A^{abc}(k^2;q^2,p^2)=:f^{abc}A(k^2;q^2,p^2)$ and $B^{abc}(k^2;q^2,p^2)=:f^{abc}B(k^2;q^2,p^2)$. 
Note, that $B$ is only relevant off-shell.

The complete DSE for the ghost-gluon vertex is derived in the
appendix\footnote{The general structure of this DSE was already given in
  \cite{Watson:priv}.}. According to the truncation scheme adopted for the
propagators \cite{vonSmekal:1997is,Fischer:2002hn}, the four-point function is neglected in the following. The truncated DSE is shown
in Fig.\ \ref{figvertex} and is given in momentum space by
\begin{eqnarray}
  \label{eq:gzDSE}
  \Gamma_{\mu}(k;q,p) &=& \Gamma_{\mu}^{(0)}(k;q,p) \nonumber \\
  &&  -\frac{1}{2}g_D^2N_c\int\frac{d^D\omega}{(2\pi)^D}\Gamma_\mu(k;\omega,\omega+k)\nonumber\\
  &&\hspace{1cm}\times  D_G(\omega)\Gamma_\nu^{(0)}(q)D_{\nu\lambda}(\omega-q)\Gamma_\lambda(q-\omega;\omega+k,p)D_G(\omega+k)\nonumber\\
   &&  -\frac{1}{2}g_D^2N_c\int\frac{d^D\omega}{(2\pi)^D}\Gamma_{\mu\nu\rho}(k,\omega,\omega+k)\nonumber\\
  &&\hspace{1cm}\times
  D_{\nu\lambda}(\omega)\Gamma_\lambda^{(0)}(q)D_G(\omega-q)\Gamma_\sigma(\omega+k;q-\omega,p)D_{\rho\sigma}(\omega+k)
  \: ,
\end{eqnarray}
where $\Gamma_\nu^{(0)}$ is the bare ghost-gluon vertex and $\Gamma_{\mu\nu\rho}$ the connected 3-gluon vertex. The momentum routing follows the same conventions as in ref.\ \cite{Alkofer:2000wg}.

Although the ghost-gluon scattering kernel is neglected, a self-consistent 
solution of this equation, together with the propagator equations, is of
significant technical complexity. Fortunately, as we will see below, such a
procedure is not necessary. It is sufficient to perform a semi-perturbative
calculation, {\it i.e.\/} to do one iteration step in the ghost-gluon vertex
DSE. If our starting hypothesis is correct, the
resulting vertex should not significantly deviate from the input tree-level
vertex. As a further test, we will also employ as input an ansatz for the
ghost-gluon vertex, which is an approximate solution of the corresponding
Slavnov-Taylor identity \cite{vonSmekal:1997is},
\begin{equation}
\Gamma^{abc}_\mu(k;q,p)=ig_Df^{abc}q_\mu \left(\frac{G(k^2)}{G(q^2)}+ \frac{G(k^2)}
{G(p^2)} -1\right)\; .
\label{GSTI}
\end{equation}

\begin{figure}
\epsfig{file=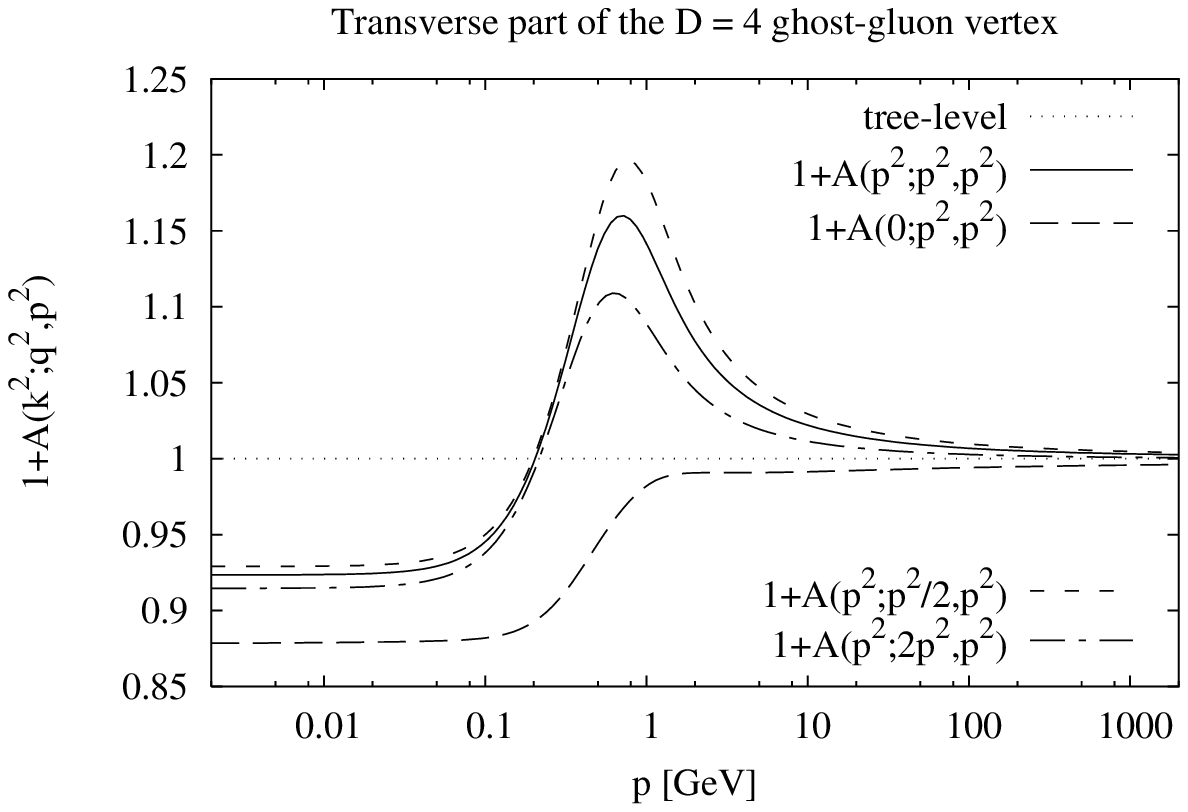,width=0.5\linewidth}\epsfig{file=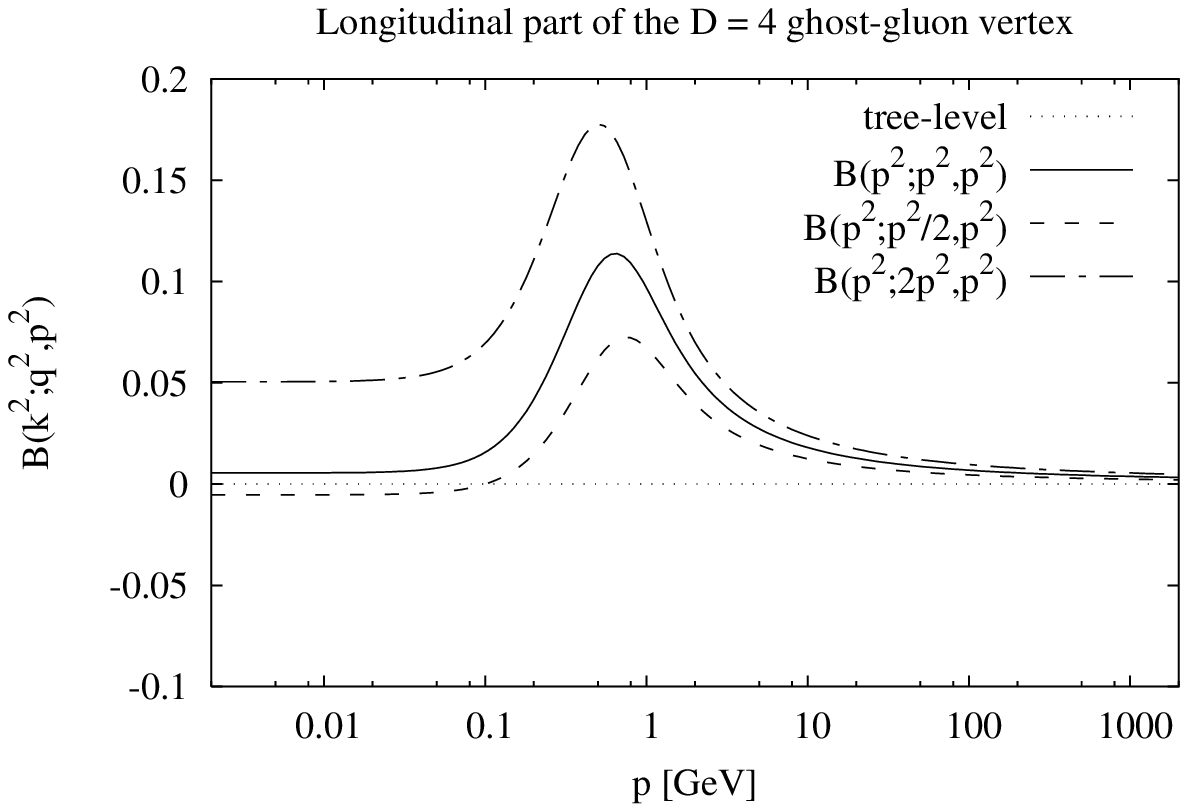,width=0.5\linewidth}
\caption{
The normalized transverse part $1+A$ %, see eq.\ (\ref{ggvtrans}),
(left panel) and the normalized longitudinal part $B$ (right panel) 
of the ghost-gluon vertex for $D=4$ in various kinematical regions.}
\label{fig4d}
\end{figure}

\begin{figure}
\epsfig{file=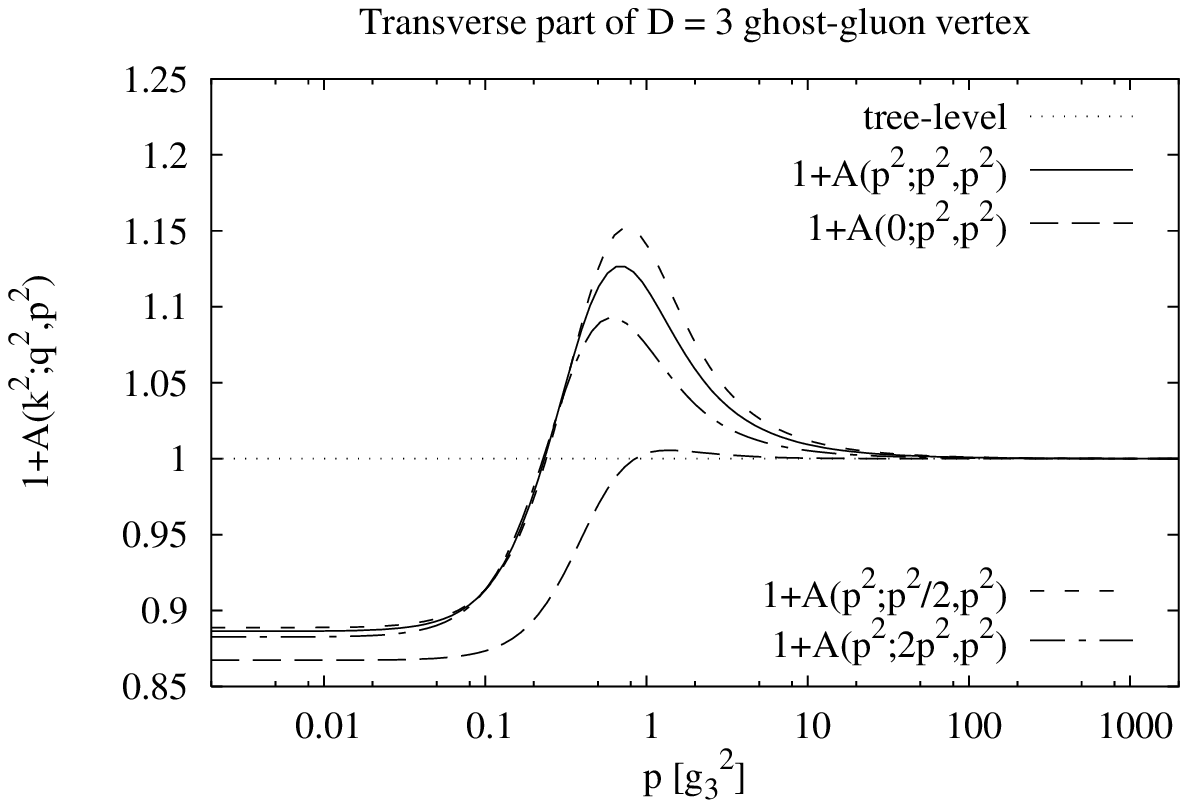,width=0.5\linewidth}\epsfig{file=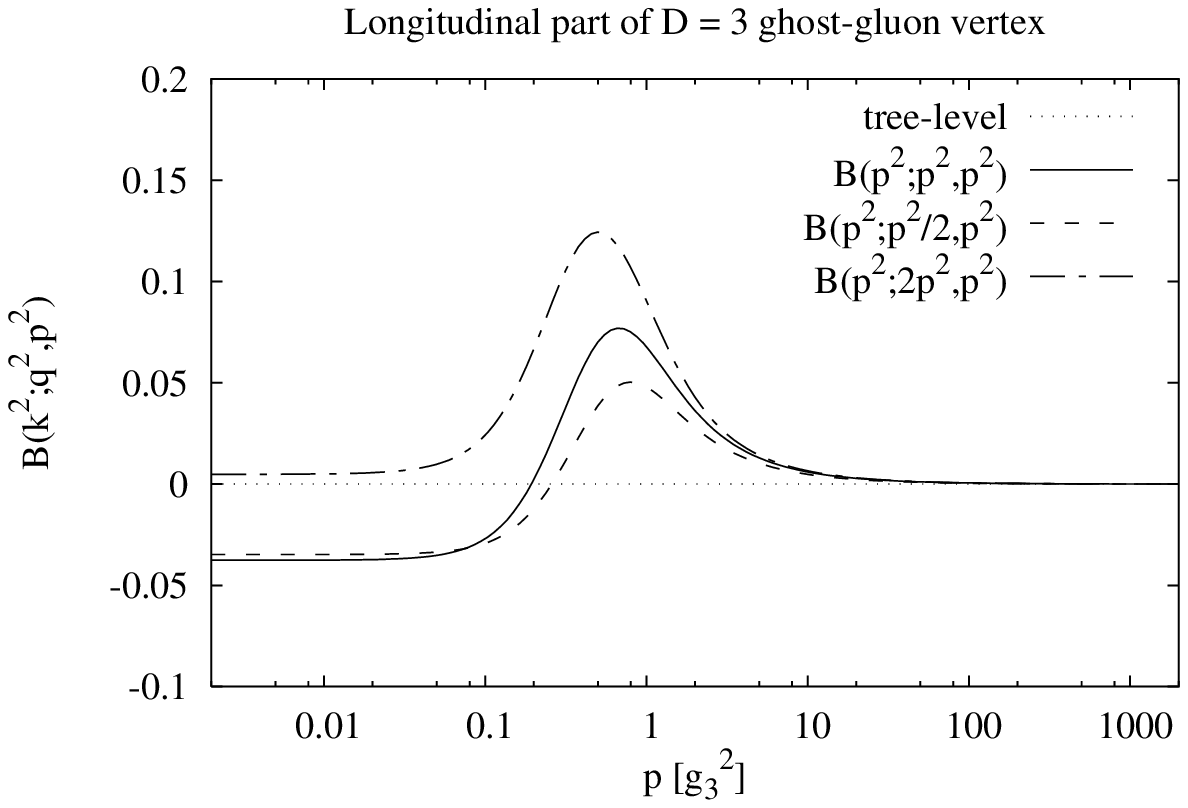,width=0.5\linewidth}
\caption{
Same as Fig.\ \ref{fig4d} for $D=3$. }
\label{fig3d}
\end{figure}

First, we will present the results with the input vertices left bare and only
the propagators dressed as described above. The results are displayed in
Fig.\ \ref{fig4d} for the $D=4$ case and in Fig.\ \ref{fig3d} for $D=3$. 
Note that these functions have the proper ghost-antighost symmetry 
\cite{Lerche:2002ep}, {\it e.g.\/} $A(k^2;q^2,p^2)=A(k^2;p^2,q^2)$ 
\cite{Schleifenbaum:diploma}. 
The transverse part of the ghost-gluon vertex, $1+A$, is extracted employing
\begin{equation}
\frac{k^2}{ig_D\Delta}q_\nu\left(\delta_{\mu\nu}-\frac{k_\mu k_\nu}{k^2}\right)
\Gamma^{abc}_\mu(k;q,p) = f^{abc} (
1+A(k^2;q^2,p^2) )\label{ggvtrans}
\end{equation}
where $\Delta = q^2 k^2-(q\cdot k)^2$ is a Gram determinant.
The deviations of the transverse part from tree-level are clearly less than
20\%. This is true for all momenta allowed by momentum conservation 
\cite{Schleifenbaum:diploma}. 
In addition, also the longitudinal part, $B(k^2;q^2,p^2)$, is smaller than
0.2 for almost all momenta and finite everywhere.

\begin{figure}
\epsfig{file=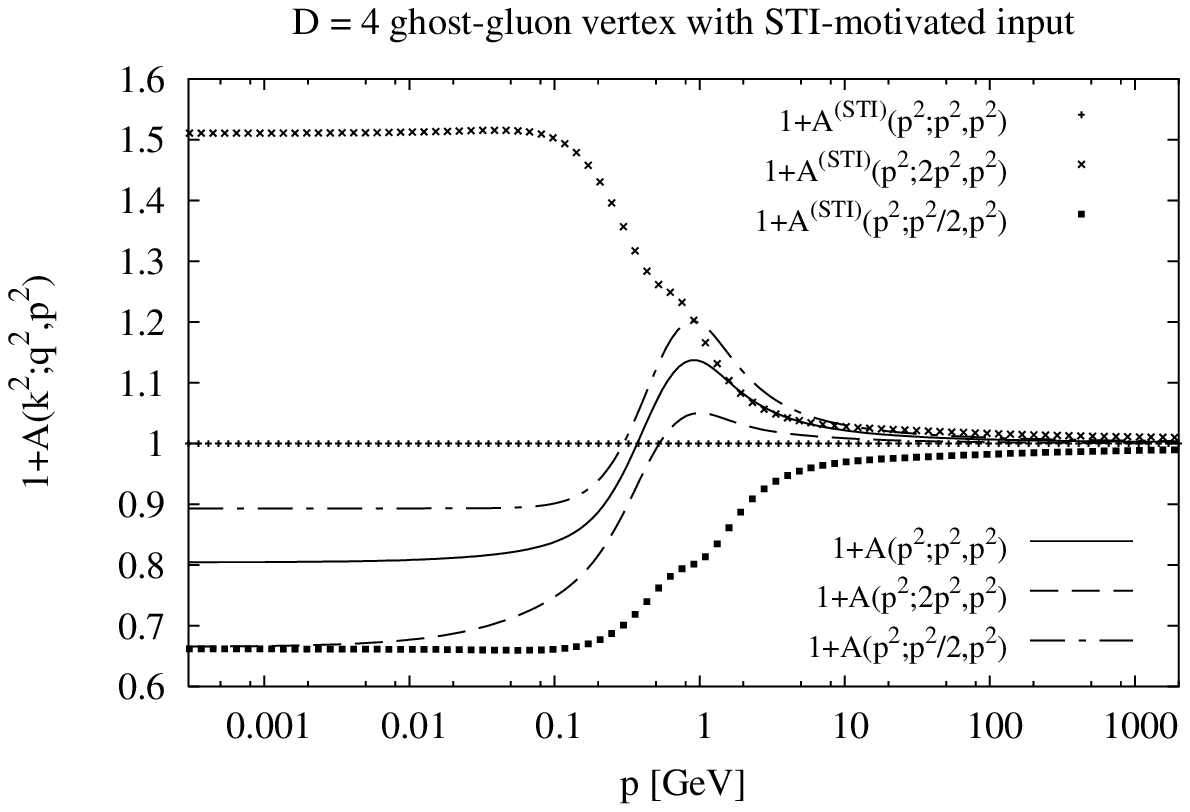,width=0.5\linewidth}\epsfig{file=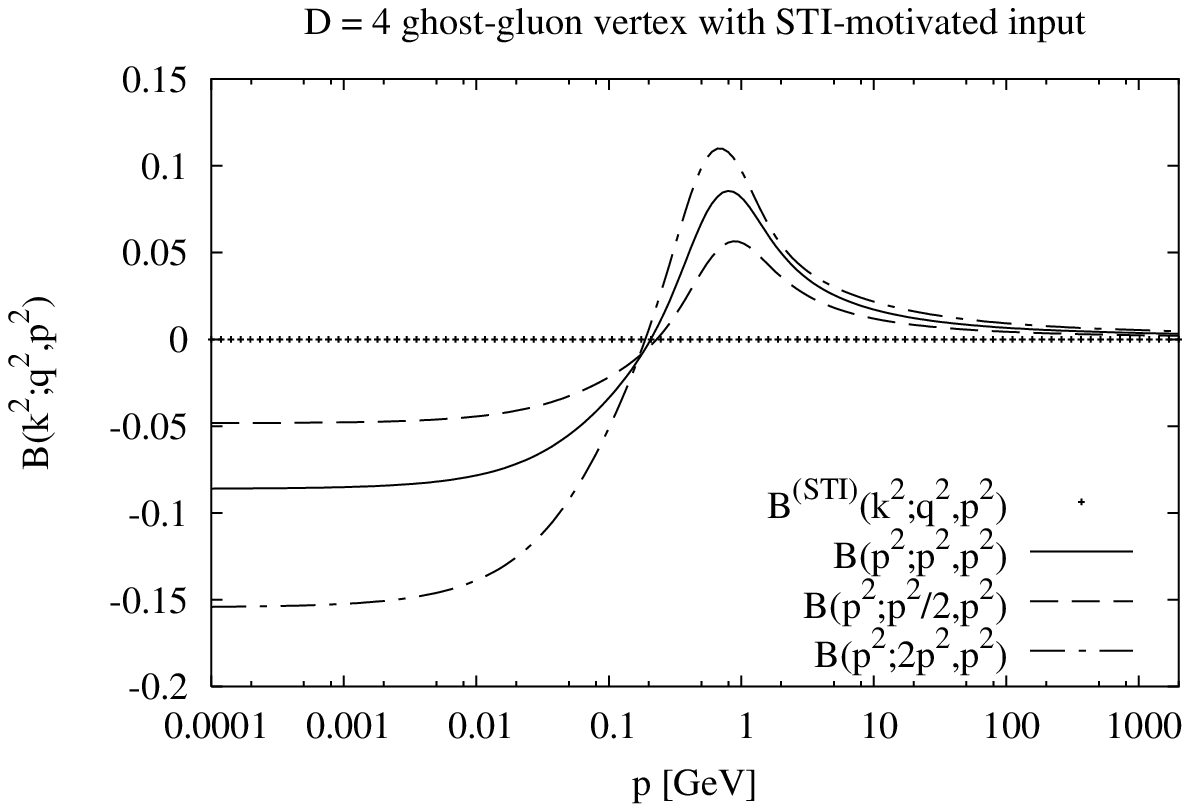,width=0.5\linewidth}
\caption{
Same as Fig.\ \ref{fig4d} but here with eq.\ (\ref{GSTI}) as input for the 
ghost-gluon vertex. The input vertices are denoted by (STI) and represented by 
crosses.}
\label{fig4dS}
\end{figure}

Thus, our results indicate that the full self-consistent solution will likely
be very close to the tree-level form. A crucial further test is provided, if 
the non-trivial form (\ref{GSTI}) is used as input on the r.h.s.\ of the 
the equation for the ghost-gluon vertex.\footnote{The 3-gluon vertex is taken bare. Recent investigations showed that the self-consistently determined
infrared
divergence of the 3-gluon vertex does not change the result presented
here \cite{Alkofer:2004it}.} Several observations can be inferred from Fig.\ \ref{fig4dS}. 
First, also in this case deviations from tree-level are small, except for those 
kinematical regions where an infrared singularity is enforced 
by the ansatz (\ref{GSTI}). Second, and even more important, the calculated vertex 
function $A$ is much closer to the tree-level case than the input. A systematic 
study of possible input vertex choices yields the same result \cite{Schleifenbaum:diploma}.
Furthermore, for $D=3$ the results are very similar 
to the $D=4$ ones \cite{Schleifenbaum:diploma}.

\begin{figure}
\epsfig{file=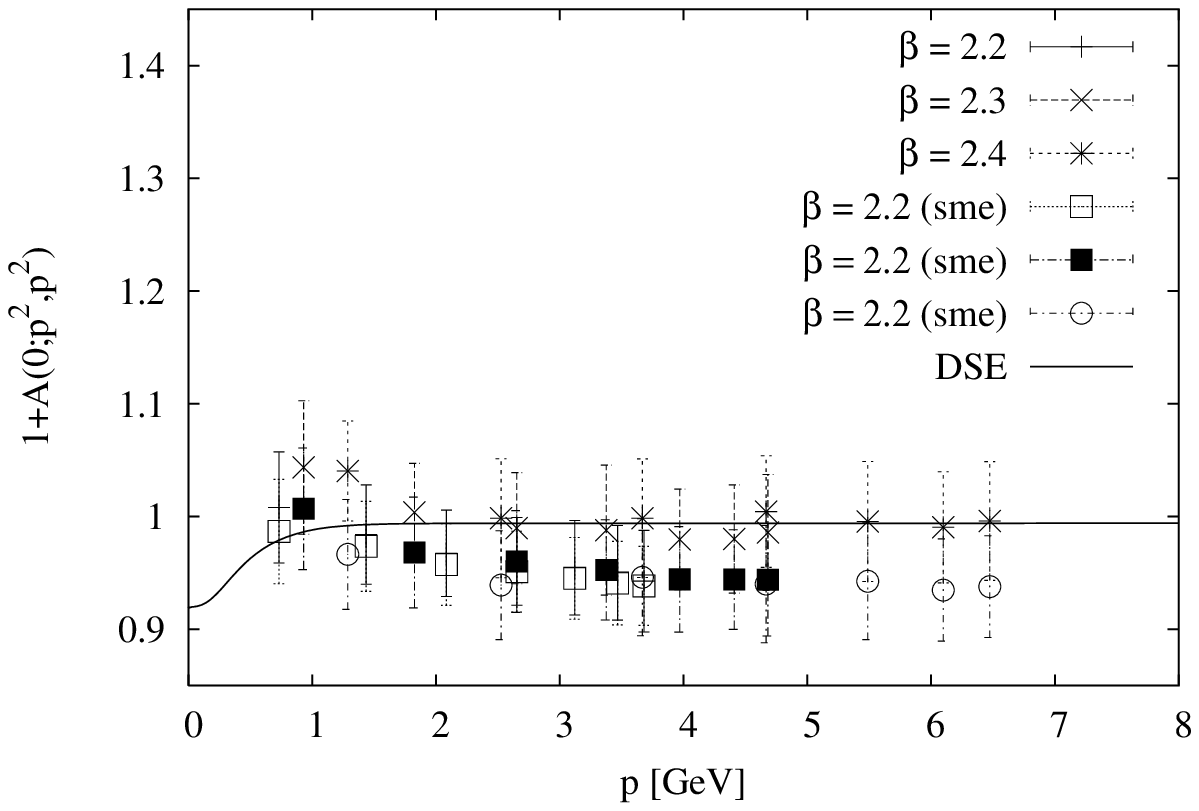,width=0.45\linewidth} \hskip 3mm
\epsfig{file=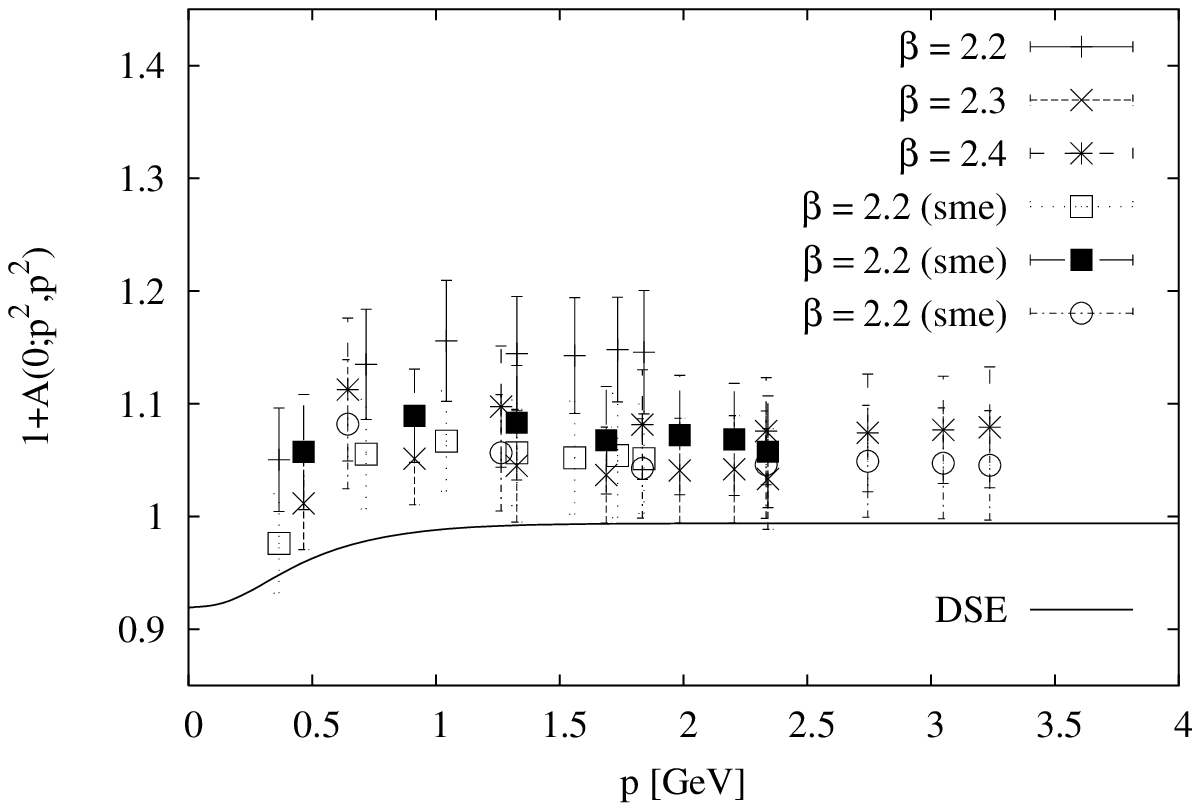,width=0.45\linewidth}
\caption{The function $1+A(0;p^2,p^2)$ for D=4 and two colors as compared 
to the corresponding lattice results \cite{Cucchieri:2004sq}. The left panel 
shows the results obtained with symmetrically chosen momenta, the right panel
with an asymmetric choice, see ref.\ \cite{Cucchieri:2004sq} for further
details.} 
\label{figLatt}
\end{figure}

Finally, we want to compare our results to recent
lattice results \cite{Cucchieri:2004sq} in Fig.\ \ref{figLatt}. 
These calculations have been performed
for gauge group SU(2). Thus, we change the color prefactor of the loop
diagrams in Fig.\ \ref{figvertex} accordingly.\footnote{This prefactor is
$f^{abc} N_c/2$ for a general number of colors.} Also, in the lattice
calculation only the ghost-gluon vertex for vanishing gluon momentum has been
determined, {\it i.e.\/} in our notation $1+A(0;p^2,p^2)$ has been calculated.
Furthermore, the smallest momentum available on the lattice is 366~MeV, and
this only at the expense of an asymmetrically chosen momentum, see ref.\ 
\cite{Cucchieri:2004sq} for more details. Note that for symmetrically chosen
momenta, see left panel of Fig.\ \ref{figLatt}, the lattice 
results for the vertex are less affected by the breaking of rotational symmetry.
Given the systematic error in
the lattice calculation, one can conclude that the lattice results are,
within errors, consistent with the tree-level form at all momenta considered.
Our results nicely match this behavior. In addition, we predict a slight 
decrease at small momenta. 

A feature of our results is the seemingly non-uniform limit for the functions
$A$ and $B$ when all three momenta vanish. This is due to the ordering of the momenta when performing the limit. As these functions are finite, the
full ghost-gluon vertex (including the corresponding prefactors, see eq.\
(\ref{tensor})) is regular. In particular, for vanishing incoming ghost momentum,
the ghost-gluon vertex is bare as expected \cite{Taylor:ff,Lerche:2002ep}.

\section{Conclusions}

We have presented approximate non-perturbative solutions for the
ghost-gluon vertex in Landau gauge for Euclidean momenta in $D=4$ and $D=3$. 
To this end, we have employed non-perturbative 
gluon and ghost propagators. We used two types of input for the ghost-gluon
vertex in the loop diagrams in order to estimate the behavior of a fully
self-consistent solution. We have also compared our results to those of a
recent lattice calculation.

These results, when taken together, show rather conclusively that deviations of 
the ghost-gluon vertex from its tree-level value are very small, 
especially in the infrared. They thus validate the truncation scheme
used to calculate the propagators of the Yang-Mills theory. More importantly,
they confirm the strong evidence for infrared ghost dominance in Landau
gauge, and thus for the Zwanziger-Gribov scenario, as they fulfill Zwanziger's
hypothesis of a bare ghost-gluon vertex in the infrared.

The results are not expected to change qualitatively when including matter fields, 
since the input propagators are rather insensitive \cite{Fischer:2003rp,Maas:2004se}
to quark contributions and no additional terms appear in the truncated DSE. This is
quite distinct from similar calculations for the quark-gluon vertex, which 
find significant deviations from the tree-level form, possibly involving
infrared divergences \cite{Llanes-Estrada:2004jz}.

In summary, the results presented here, nicely match a picture of confinement 
where the confining fields are on or near the Gribov horizon. They provide a
further piece of evidence for a confinement mechanism of the Kugo-Ojima or
Zwanziger-Gribov type.

\section*{Acknowledgments}
We thank Attilio Cucchieri, Christian S.\ Fischer, Tereza Mendes, Lorenz von
Smekal, Peter Watson and Daniel Zwanziger for many valuable  discussions and
comments on the manuscript. We are especially grateful to Attilio Cucchieri 
and Tereza Mendes for providing and explaining their lattice results. We
furthermore thank Peter Watson for communicating his derivation of the
ghost-gluon vertex DSE.

This work was supported by the BMBF under grant number 06DA917,
and by the Helmholtz association (Virtual Theory Institute VH-VI-041).

\appendix
\renewcommand{\theequation}{A\arabic{equation}}
  % redefine the command that creates the equation no.
\setcounter{equation}{0}  % reset counter 
\section*{Appendix: Derivation of the DSE for the ghost-gluon vertex}
\label{sec:DSEcalc}
To derive the DSE for the ghost-gluon vertex it is only necessary to consider an action
${\cal S}_{gh}$ that involves the contributions from ghosts in the Lagrangian (\ref{lagrange}), although due to the mutual
coupling of the Green functions, the entire Lagrangian is implicitly relevant. Furthermore, we introduce $J_\mu^a$, $\bar{\sigma}^a$ and
$\sigma^a$ as sources for the fields $A_\mu^a$, $c^a$ and $\bar{c}^a$,
respectively, so that we can define the generating functional for full Green
functions,
\begin{equation}
  Z[J,\bar{\sigma},\sigma] = \int{\cal D}[A \bar{c} c]\exp\left(-\int d^Dx{\cal L}+\int d^D x \left(A^a_\mu J^a_\mu + \bar{\sigma}^a c^a + \bar{c}^a \sigma^a \right) \right),
\end{equation}
the generating functional for connected Green functions,
$W[J,\bar{\sigma},\sigma]=\ln Z$, and the one for proper Green functions,
$\Gamma[A,c,\bar{c}]=-W[J,\bar{\sigma},\sigma]+\int d^dx (A^a_\mu J^a_\mu + \bar{\sigma}^a c^a + \bar{c}^a \sigma^a)$. The fields and sources
are then given by
\begin{eqnarray}
  \label{eq:deriv}
  \frac{\delta W}{\delta \sigma^a} &=& \bar{c}^a, \hspace*{1cm}\frac{\delta W}{\delta \bar{\sigma}^a} = c^a, 
\hspace*{1cm}\frac{\delta W}{\delta J_\mu^a} = A_\mu^a, \nonumber\\
\frac{\delta \Gamma}{\delta c^a} &=& \bar{\sigma}^a, \hspace*{1cm}
\frac{\delta \Gamma}{\delta \bar{c}^a} = {\sigma}^a, 
\hspace*{1cm}\frac{\delta \Gamma}{\delta A_\mu^a} = J_\mu^a ,
\end{eqnarray}
where we use right derivatives for the following Grassmann fields,
\begin{equation}
  \frac{\delta}{\delta c^a} :=
  \frac{\overleftarrow{\delta}}{\delta c^a}\quad ,
  \quad\quad\frac{\delta}{\delta \si^a}
  := \frac{\overleftarrow{\delta}}{\delta \si^a}\: .
\end{equation}
One way to approach the DSE for the ghost-gluon vertex is to start with the identity
\begin{eqnarray}
  0 &=& \int{\cal D}[A \bar{c} c]\frac{\delta}{\delta \bar{c}^b(y)}\exp\left(-\int d^D x{\cal L}+\int d^D x \left(A^a_\mu J^a_\mu + \bar{\sigma}^a c^a + \bar{c}^a \sigma^a  \right) \right) \nonumber \\
&=& \left\langle -\frac{\delta {\cal S}_{gh}\left[A,c,\bar{c}\right]}{\delta \bar{c}^b(y)} +
  \sigma^b(y) \right\rangle\label{eq:basedse}
\: .
\label{eq:DSEAPPstart}
\end{eqnarray}
The expression in the brackets represents a full correlation function as
generated by $Z$. Retaining non-zero sources, we now apply to the above
expression the derivative
\begin{equation}
  \frac{\delta}{\delta c^c(z)}=\int d^D v \frac{\delta^2 \Gamma}{ \delta \ovc^d(v)\delta c^c(z)
 } \frac{\delta}{\delta \si^d(v)}+ \mbox{vanishing terms} \: .
\end{equation}
Some terms vanish because the functionals can depend on pairs
of Grassmann fields only. Throughout the calculation, great care is mandatory when dropping terms since most of
them vanish only when setting sources to zero. 
We define the ghost and gluon propagators in position space
\begin{eqnarray}
  \label{eq:defProps}
  \tilde{D}_G^{ab}(x-y) & := & \left.\left<c^a(x)\ovc^b(y)\right>\right|_{\eta\equiv 0}=\left.\frac{\delta^2 W}{\delta \ovs^a(x) \delta
  \si^b(y)}\right|_{\eta\equiv 0}\: , \\
  \tilde{D}_{\mu\nu}^{ab}(x-y) & := & \left.\left<A^a_\mu(x)A^b_\nu(y)\right>\right|_{\eta\equiv 0}=\left.\frac{\delta^2 W}{\delta
  J_{\mu}^a(x) \delta J_{\nu}^b(y)}\right|_{\eta\equiv 0} \: ,
\end{eqnarray}
as well as the proper ghost-gluon vertex in position space
\begin{eqnarray}
  \label{eq:defGZ}
  \oG^{abc}_{\mu}(x;y,z):=\left.\frac{\delta^3 \Gamma}{\delta A_{\mu}^a(x)
  \delta \ovc^b(y) \delta c^c(z)}\right|_{\eta\equiv 0} \: .
\end{eqnarray}
After usage of the relation
\begin{eqnarray}
\label{eq:id}
  \delta(x-y)\delta^{ab} = \frac{\delta \bar{\sigma}^b(y)}{\delta \bar{\sigma}^a(x)}
=\int d^Dz \frac{\delta \bar{\sigma}^b(y)}{\delta \bar{c}^d(z)}\frac{\delta \bar{c}^d(z)}{\delta \bar{\sigma}^a(x)}
&=&\int d^Dz \frac{\delta^2 \Gamma}{\delta \bar{c}^d(z) \delta c^b(y)} \frac{\delta^2 W}
{\delta \bar{\sigma}^a(x) \delta \sigma^d(z)} \nonumber\\
\end{eqnarray}
one then obtains
\begin{eqnarray}
  \frac{\delta^2 \Gamma}{\delta \ovc^b(y)\delta c^c(z)}Z[J,\ovs,\si]  &=&
 \partial^{2}\delta^{bc}\delta(y-z)Z[J,\ovs,\si] \nonumber \\ &&+ g_Df^{bgh}\partial_{\rho}^y\int d^D v \frac{\delta^2 \Gamma}{ \delta \ovc^d(v)\delta c^c(z)
 }\left\langle A_{\rho}^h(y) c^g(y)\ovc^d(v)\right\rangle\: .
\label{eq:Ghost2}
\end{eqnarray}
To find the DSE for the ghost-gluon vertex, we apply to equation (\ref{eq:Ghost2}) the derivative
\begin{equation}
  \frac{\delta}{\delta A^a_{\mu}(x)}=\int d^D u \frac{\delta^2 \Gamma}{\delta A^a_{\mu}(x)
  \delta A^e_{\nu}(u)} \frac{\delta}{\delta J^e_{\nu}(u)}+ \mbox{vanishing
  terms}\: .
\end{equation}
One can now immediately set sources to zero to find the proper ghost-gluon vertex:
\begin{eqnarray}
\oG^{abc}_{\mu}(x;y,z) &=& g_Df^{bgh}\partial_{\rho}^y\int d^D v \left.\frac{\delta^3 \Gamma}{\delta A^a_{\mu}(x) \delta \ovc^d(v)\delta c^c(z)
 }\left\langle A_{\rho}^h(y) c^g(y)\ovc^d(v)\right\rangle\right|_{\eta\equiv 0}\nonumber\\
   && \hspace*{-2.7cm}+ g_Df^{bgh}\partial_{\rho}^y\int d^D [uv] \left.\frac{\delta^2 \Gamma}{ \delta \ovc^d(v)\delta c^c(z)
 }\frac{\delta^2 \Gamma}{\delta A^a_{\mu}(x)
  \delta A^e_{\nu}(u)}\left\langle A_{\rho}^h(y) A^e_{\nu}(u) 
  c^g(y)\ovc^d(v)\right\rangle\right|_{\eta\equiv 0} \: .
\label{eq:GZvert1}
\end{eqnarray}
The decomposition of the full 4-point
correlation function yields
\begin{eqnarray}
  \label{eq:dec4pt}
  \left.\left\langle A_{\rho}^h(y) A^e_{\nu}(u) 
  c^g(y)\ovc^d(v)\right\rangle\right|_{\eta\equiv 0} &&\nonumber \\
  &&\hspace*{-5.2cm}= \quad\left.\frac{\delta^2
  W}{ \delta J_{\rho}^h(y) \delta J^e_{\nu}(u)}\frac{\delta^2
  W}{\delta \ovs^g(y)\delta \si^d(v)}\right|_{\eta\equiv 0}+ 
  \left.\frac{\delta}{\delta J_{\rho}^h(y)}\frac{\delta^3
  W}{\delta J^e_{\nu}(u) 
  \delta \ovs^g(y) \delta \si^d(v)}\right|_{\eta\equiv 0} \nonumber\\
  &&\hspace*{-5.2cm}= \left.\quad\frac{\delta^2
  W}{\delta J_{\rho}^h(y) \delta J^e_{\nu}(u)}\frac{\delta^2
  W}{\delta \ovs^g(y)\delta \si^d(v)}\right|_{\eta\equiv 0}+\int d^D [rst]\times \nonumber \\
  &&\hspace*{-4.9cm} \left\{ -
  \frac{\delta^2 W}{\delta J^e_{\nu}(u) \delta J_{\lambda}^k(r)}\frac{\delta^2 W}{\delta \ovs^g(y) \delta \si^m(s)}\frac{\delta^3
  \Gamma}{\delta A_{\lambda}^k(r) \delta \ovc^m(s) \delta c^n(t)}\frac{\delta^3
  W}{\delta J_{\rho}^h(y) \delta \ovs^n(t)
  \delta \si^d(v)} \right. \nonumber \\
  &&\hspace*{-4.6cm} -
  \frac{\delta^3 W}{\delta J_{\rho}^h(y) \delta J^e_{\nu}(u) \delta J_{\lambda}^k(r)}\frac{\delta^2 W}{\delta \ovs^g(y) \delta \si^m(s)}\frac{\delta^3
  \Gamma}{\delta A_{\lambda}^k(r) \delta \ovc^m(s) \delta c^n(t)}\frac{\delta^2 W}{\delta \ovs^n(t)
  \delta \si^d(v)} \nonumber \\
  &&\hspace*{-4.6cm} -
  \int d^Dw  \frac{\delta^2 W}{\delta J^e_{\nu}(u) \delta J_{\lambda}^k(r)}\frac{\delta^2 W}{\delta J^h_{\rho}(y) \delta J_{\sigma}^l(w)}\frac{\delta^2 W}{\delta \ovs^g(y) \delta \si^m(s)}\nonumber\\&&\hspace*{-3.0cm}\times\frac{\delta^4
  \Gamma}{\delta A_{\sigma}^l(w) \delta A_{\lambda}^k(r) \delta \ovc^m(s) \delta c^n(t)}\frac{\delta^2 W}{\delta \ovs^n(t)
  \delta \si^d(v)}  \\
  &&\hspace*{-4.6cm} \left.\left. -
  \frac{\delta^2 W}{\delta J^e_{\nu}(u) \delta
  J_{\lambda}^k(r)}\frac{\delta^3 W}{\delta J_{\rho}^h(y) \delta \ovs^g(y) \delta \si^m(s)}\frac{\delta^3
  \Gamma}{\delta A_{\lambda}^k(r) \delta \ovc^m(s) \delta c^n(t)}\frac{\delta^2 W}{\delta \ovs^n(t)
  \delta \si^d(v)}\right\}\right|_{\eta\equiv 0}  .\nonumber
\end{eqnarray}
The last term of the last line in eq.\ (\ref{eq:dec4pt}) produces a 3PI-graph which, however, cancels in
eq.\ (\ref{eq:GZvert1}) with the first term. We now introduce two further
definitions. The proper three-gluon vertex shall be denoted by
\begin{equation}
  \label{eq:def3Z}
  \oG^{and}_{\mu\nu\lambda}(x,w,r):=\left.\frac{\delta^3 \Gamma}{\delta
  A_{\mu}^a(x) \delta A^n_{\nu}(w) \delta A_{\lambda}^d(r)}\right|_{\eta\equiv
  0} \: ,
\end{equation}
and the proper 4-point Green function involving two gluons and two ghosts is defined by
\begin{equation}
  \label{eq:def2G2Z}
  \oG^{nagc}_{\sigma\mu}(w,x;t,z):=\left.\frac{\delta^4
  \Gamma}{\delta A_{\sigma}^n(w) \delta A_{\mu}^a(x) \delta \ovc^g(t) \delta c^c(z)}\right|_{\eta\equiv 0} \: .
\end{equation}
After further decompositions of connected into proper 3-point correlation functions
 and using (\ref{eq:id}),
eq.\ (\ref{eq:GZvert1}) can be rewritten\footnote{The indices and integration
  variables have been renamed in a convenient way.} as
\begin{eqnarray}
  \label{eq:GZvert2}
  \oG^{abc}_{\mu}(x;y,z) &=& g_Df^{abc}\partial_{\mu}^y \delta(y-x)\delta(y-z)
    \nonumber\\
  && \hspace*{-2.6cm}+  g_Df^{hbm}\partial_{\rho}^y \int d^D[rstw]\tilde{D}_{\rho\sigma}^{hg}(y-t)\tilde{D}_G^{mn}(y-w) \oG_{\mu}^{and}(x;w,r) 
  \tilde{D}_G^{de}(r-s)\oG_{\sigma}^{gec}(t;s,z) \nonumber \\
  && \hspace*{-2.6cm}+  g_Df^{mbh}\partial_{\rho}^y \int d^D[rstw]\tilde{D}_G^{hg}(y-t) \tilde{D}_{\rho\nu}^{mn}(y-w) \oG_{\mu\nu\lambda}^{and}(x,w,r)
  \tilde{D}_{\lambda\sigma}^{de}(r-s)\oG_{\sigma}^{egc}(s;t,z)
  \nonumber\\
  && \hspace*{-2.6cm}- g_Df^{mbh}\partial_{\rho}^y\int d^D[tw] \tilde{D}_G^{hg}(y-t) \oG_{\sigma\mu}^{nagc}(w,x;t,z)
  \tilde{D}_{\rho\sigma}^{mn}(y-w) \: .
\end{eqnarray}
The last step to take is to identify the bare ghost-gluon vertex which is
derived from the Lagrangian (\ref{lagrange}) as
\begin{equation}
  \label{eq:defbareGZ}
  \oG^{(0)abc}_{\mu}(x;y,z):=\frac{\delta^3 \mathcal{S}_{gh}}{\delta A_{\mu}^a(x)
  \delta \ovc^b(y) \delta c^c(z)}=g_Df^{abc}\partial_{\mu}^y
  \delta(y-x)\delta(y-z) \: .
\end{equation}
Thus one readily obtains
\begin{equation}
  \label{eq:partials}
  g_Df^{hbm}\partial_{\rho}^y \tilde{D}_{\rho\sigma}^{hg}(y-t)
  \tilde{D}_G^{mn}(y-w) = \int d^d[uv]
  \oG_{\rho}^{(0)hbm}(u;y,v)\tilde{D}_{\rho\sigma}^{hg}(u-t)
  \tilde{D}_G^{mn}(v-w)  \: .
\end{equation}
Using this, one can remove the spacetime derivatives in favor of
bare ghost-gluon vertices and finally arrive at the complete DSE for the ghost-gluon vertex in position space:
\begin{eqnarray}
  \label{eq:GZdsePOS}
  \oG^{abc}_{\mu}(x;y,z) &=& \oG^{(0)abc}_{\mu}(x;y,z)
    \nonumber\\
  &&\hspace*{-2.6cm} +  \int d^d[rstuvw]\tilde{D}_G^{mn}(v-w) \oG_{\mu}^{and}(x;w,r) 
  \tilde{D}_G^{de}(r-s)\oG_{\sigma}^{gec}(t;s,z)\tilde{D}_{\rho\sigma}^{hg}(u-t)\oG^{(0)hbm}_{\mu}(u;y,v) \nonumber \\
  &&\hspace*{-2.6cm} + \int d^d[rstuvw]\tilde{D}_{\rho\nu}^{mn}(u-w) \oG_{\mu\nu\lambda}^{and}(x,w,r)
  \tilde{D}_{\lambda\sigma}^{de}(r-s) \oG_{\sigma}^{egc}(s;t,z)\tilde{D}_G^{hg}(v-t)\oG_{\rho}^{(0)mbh}(u;y,v)
  \nonumber\\
  &&\hspace*{-2.6cm} - \int d^d[tuvw] \tilde{D}_G^{gh}(v-t) \oG_{\sigma\mu}^{nagc}(w,x;t,z)
  \tilde{D}_{\rho\sigma}^{mn}(u-w)\oG^{(0)mbh}_{\mu}(u;y,v) \: .
\end{eqnarray}
From this equation, eq.\ (\ref{eq:gzDSE}) follows straightforwardly by
Fourier
transformation.

%\newpage

{%\small 
}


\begin{thebibliography}{99}

%\parskip=0pt

\bibitem{Maris:2003vk}
P.~Maris and C.~D.~Roberts,
%``Dyson-Schwinger equations: A tool for hadron physics,''
Int.\ J.\ Mod.\ Phys.\ E {\bf 12} (2003) 297
[arXiv:nucl-th/0301049].
%%CITATION = NUCL-TH 0301049;%%

\bibitem{Alkofer:2000wg}
R.~Alkofer and L.~von Smekal,
%``The infrared behavior of QCD Green's functions: 
%Confinement, dynamical  symmetry breaking, and hadrons as relativistic 
%bound states,''
Phys.\ Rept.\  {\bf 353} (2001) 281 
[arXiv:hep-ph/0007355].
%%CITATION = HEP-PH 0007355;%%

\bibitem{Roberts:2000aa}
C.~D.~Roberts and S.~M.~Schmidt,
Prog.\ Part.\ Nucl.\ Phys.\ {\bf 45} (2000) S1 
[arXiv:nucl-th/0005064].
%%CITATION = NUCL-TH 0005064;%%

\bibitem{Fischer:2003rp}
C.~S.~Fischer and R.~Alkofer,
%``Non-perturbative propagators, running coupling and dynamical quark mass  of Landau gauge QCD,''
Phys.\ Rev.\ D {\bf 67} (2003) 094020
[arXiv:hep-ph/0301094].
%%CITATION = HEP-PH 0301094;%%

\bibitem{Alkofer:2003jj}
R.~Alkofer, W.~Detmold, C.~S.~Fischer and P.~Maris,
%``Analytic properties of the Landau gauge gluon and quark propagators,''
Phys.\ Rev.\ D {\bf 70} (2004) 014014 [arXiv:hep-ph/0309077].
%%CITATION = HEP-PH 0309077;%%

\bibitem{Pawlowski:2003hq}
J.~M.~Pawlowski, D.~F.~Litim, S.~Nedelko and L.~von Smekal,
%``Infrared behaviour and fixed points in Landau gauge QCD,''
Phys.\ Rev.\ Lett.\  {\bf 93} (2004) 152002 [arXiv:hep-th/0312324];
%%CITATION = HEP-TH 0312324;%%
%\cite{Fischer:2004uk}
%\bibitem{Fischer:2004uk}
C.~S.~Fischer and H.~Gies,
%``Renormalization flow of Yang-Mills propagators,''
arXiv:hep-ph/0408089.
%%CITATION = HEP-PH 0408089;%%

\bibitem{Zwanziger:2003cf}
D.~Zwanziger,
%``Non-perturbative Faddeev-Popov formula and infrared limit of QCD,''
Phys.\ Rev.\ D {\bf 69} (2004) 016002 [arXiv:hep-ph/0303028].
%%CITATION = HEP-PH 0303028;%%

\bibitem{Sternbeck:2004xr}
A.~Sternbeck, E.~M.~Ilgenfritz, M.~Muller-Preussker and A.~Schiller,
%``The gluon and ghost propagator and the influence of Gribov copies,''
arXiv:hep-lat/0409125.
%%CITATION = HEP-LAT 0409125;%%

\bibitem{Cucchieri:2004mf}
A.~Cucchieri, T.~Mendes and A.~R.~Taurines,
%``Positivity violation for the lattice Landau gluon propagator,''
Phys.\ Rev.\ D {\bf 71}, 051902 (2005) [arXiv:hep-lat/0406020].
%%CITATION = HEP-LAT 0406020;%%

\bibitem{Oliveira:2004gy}
O.~Oliveira and P.~J.~Silva,
%``The infrared Landau gauge gluon propagator from lattice QCD,''
AIP Conf.\ Proc.\  {\bf 756} (2005) 290
[arXiv:hep-lat/0410048].
%%CITATION = HEP-LAT 0410048;%%

\bibitem{Bowman:2004jm}
P.~O.~Bowman, U.~M.~Heller, D.~B.~Leinweber, M.~B.~Parappilly and A.~G.~Williams,
%``Unquenched gluon propagator in Landau gauge,''
Phys.\ Rev.\ D {\bf 70} (2004) 034509
[arXiv:hep-lat/0402032].
%%CITATION = HEP-LAT 0402032;%%

\bibitem{Gattnar:2004bf}
J.~Gattnar, K.~Langfeld and H.~Reinhardt,
%``Signals of confinement in Green functions of SU(2) Yang-Mills theory,''
Phys.\ Rev.\ Lett.\  {\bf 93} (2004) 061601
[arXiv:hep-lat/0403011].
%%CITATION = HEP-LAT 0403011;%%
\bibitem{Zwanziger:1993dh}
D.~Zwanziger,
%``Fundamental modular region, Boltzmann factor and area law in lattice gauge theory,''
Nucl.\ Phys.\ B {\bf 412} (1994) 657.
%%CITATION = NUPHA,B412,657;%%

\bibitem{Gribov:1977wm}
V.~N.~Gribov,
%``Quantization Of Non-Abelian Gauge Theories,''
Nucl.\ Phys.\ B {\bf 139} (1978) 1.
%%CITATION = NUPHA,B139,1;%%

\bibitem{Kugo:1995km}
T.~Kugo,
%``The Universal Renormalization Factors Z_1/Z_3 and Color Confinement Condition in Non-Abelian Gauge Theory,''
arXiv:hep-th/9511033.
%%CITATION = HEP-TH 9511033;%%

\bibitem{Kugo:gm}
T.~Kugo and I.~Ojima,
%``Local Covariant Operator Formalism Of Nonabelian Gauge Theories And Quark Confinement Problem,''
Prog.\ Theor.\ Phys.\ Suppl.\  {\bf 66} (1979) 1.
%%CITATION = PTPSA,66,1;%%

\bibitem{Oehme:bj}
R.~Oehme and W.~Zimmermann,
%``Gauge Field Propagator And The Number Of Fermion Fields,''
Phys.\ Rev.\ D {\bf 21} (1980) 1661;
%%CITATION = PHRVA,D21,1661;%%
%R.~Oehme and W.~Zimmermann,
%``Quark And Gluon Propagators In Quantum Chromodynamics,''
Phys.\ Rev.\ D {\bf 21} (1980) 471.
%%CITATION = PHRVA,D21,471;%%

\bibitem{Feuchter:2004gb}
C.~Feuchter and H.~Reinhardt,
%``Quark and gluon confinement in Coulomb gauge,''
arXiv:hep-th/0402106;
%%CITATION = HEP-TH 0402106;%%
%\bibitem{Feuchter:2004mk}
arXiv:hep-th/0408236;
%%CITATION = HEP-TH 0408236;%%
%\bibitem{Reinhardt:2004mm}
H.~Reinhardt and C.~Feuchter,
%``On the Yang-Mills wave functional in Coulomb gauge,''
arXiv:hep-th/0408237;
%%CITATION = HEP-TH 0408237;%%
%\cite{Zwanziger:2003de}
D.~Zwanziger,
%``Analytic calculation of color-Coulomb potential and color confinement,''
Phys.\ Rev.\ D {\bf 70} (2004) 094034
[arXiv:hep-ph/0312254];
%%CITATION = HEP-PH 0312254;%%
A.~P.~Szczepaniak and E.~S.~Swanson,
%``Coulomb gauge QCD, confinement, and the constituent representation,''
Phys.\ Rev.\ D {\bf 65} (2002) 025012
[arXiv:hep-ph/0107078].
%%CITATION = HEP-PH 0107078;%%

\bibitem{Greensite:2004ur}
J.~Greensite, S.~Olejnik and D.~Zwanziger,
%``Center vortices and the Gribov horizon,''
arXiv:hep-lat/0407032.
%%CITATION = HEP-LAT 0407032;%%

\bibitem{Taylor:ff}
J.~C.~Taylor,
%``Ward Identities And Charge Renormalization Of The Yang-Mills Field,''
Nucl.\ Phys.\ B {\bf 33} (1971) 436;
%%CITATION = NUPHA,B33,436;%%
W.~J.~Marciano and H.~Pagels,
%``Quantum Chromodynamics: A Review,''
Phys.\ Rept.\  {\bf 36} (1978) 137.
%%CITATION = PRPLC,36,137;%%

\bibitem{Lerche:2002ep}
C.~Lerche and L.~von Smekal,
%``On the infrared exponent for gluon and ghost propagation in Landau  gauge QCD,''
Phys.\ Rev.\ D {\bf 65} (2002) 125006
[arXiv:hep-ph/0202194].
%%CITATION = HEP-PH 0202194;%%

\bibitem{vonSmekal:1997is}
L.~von Smekal, R.~Alkofer and A.~Hauck,
%``The infrared behavior of gluon and ghost propagators in Landau gauge  QCD,''
Phys.\ Rev.\ Lett.\  {\bf 79} (1997) 3591 %\newline
[arXiv:hep-ph/9705242];
%%CITATION = HEP-PH 9705242;%%
%\bibitem{vonSmekal:1998is}
%L.~von Smekal, A.~Hauck and R.~Alkofer,
%``A solution to coupled Dyson-Schwinger equations for gluons and ghosts  
%in Landau gauge,''
Annals Phys.\  {\bf 267} (1998) 1
%[Erratum-ibid.\  {\bf 269}, 182 (1998)]
[arXiv:hep-ph/9707327].
%%CITATION = HEP-PH 9707327;%%

%\cite{Ellwanger:1996wy}
\bibitem{Ellwanger:1996wy}
U.~Ellwanger, M.~Hirsch and A.~Weber,
%``The heavy quark potential from Wilson's exact renormalization group,''
Eur.\ Phys.\ J.\ C {\bf 1} (1998) 563
[arXiv:hep-ph/9606468].
%%CITATION = HEP-PH 9606468;%%

\bibitem{Cucchieri:2004sq}
A.~Cucchieri, T.~Mendes and A.~Mihara,
%``Numerical study of the ghost-gluon vertex in Landau gauge,''
JHEP {\bf 0412}, 012 (2004) [arXiv:hep-lat/0408034].
%%CITATION = HEP-LAT 0408034;%%

\bibitem{Fischer:2002hn}
C.~S.~Fischer  and R.~Alkofer,
%{\it Infrared Exponents and Running Coupling of SU(N) Yang--Mills Theories},
Phys. Lett. B {\bf 536} (2002) 177 
[arXiv:hep-ph/0202202];
%%CITATION = HEP-PH 0202202;%%
%\cite{Fischer:2002eq}
%\bibitem{Fischer:2002eq}
C.~S.~Fischer, R.~Alkofer and H.~Reinhardt,
%{\it The Elusiveness of Infrared Critical Exponents in Landau
%Gauge Yang-Mills Theories},
Phys.\ Rev.\ D {\bf 65}  (2002) 094008
[arXiv: hep-ph/0202195].
%%CITATION = HEP-PH 0202195;%%

\bibitem{Maas:2004se}
A.~Maas, J.~Wambach, B.~Gr\"uter and R.~Alkofer,
%``High-temperature limit of Landau-gauge Yang-Mills theory,''
Eur. Phys. J. {\bf C37} (2004) 335
[arXiv:hep-ph/0408074].
%%CITATION = HEP-PH 0408074;%%

\bibitem{Fischer:2004ym}
C.~S.~Fischer, F.~Llanes-Estrada and R.~Alkofer,
%``Dynamical mass generation in Landau gauge QCD,''
arXiv:hep-ph/0407294.
%%CITATION = HEP-PH 0407294;%%

\bibitem{Maas:2004ec}
A.~Maas, J.~Wambach, B.~Gr\"uter and R.~Alkofer,
%``Residual confinement in high-temperature Yang-Mills theory,''
arXiv:hep-ph/0408299.
%%CITATION = HEP-PH 0408299;%%

\bibitem{Zwanziger:2001kw}
D.~Zwanziger,
%``Non-perturbative Landau gauge and infrared critical exponents in QCD,''
Phys.\ Rev.\ D {\bf 65}  (2002) 094039
[arXiv: hep-th/0109224].
%%CITATION = HEP-TH 0109224;%%

\bibitem{Watson:2001yv}
P.~Watson and R.~Alkofer,
%``Verifying the Kugo-Ojima confinement criterion in Landau gauge QCD,''
Phys.\ Rev.\ Lett.\  {\bf 86} (2001) 5239 
[arXiv:hep-ph/0102332].
%%CITATION = HEP-PH 0102332;%%

\bibitem{Schleifenbaum:diploma}
W.~Schleifenbaum, diploma thesis, TU Darmstadt, 2004.

\bibitem{Watson:priv}
P.~Watson, private communication; internal report (T\"ubingen university).

\bibitem{Alkofer:2004it}
R.~Alkofer, C.~S.~Fischer and F.~J.~Llanes-Estrada,
%``Vertex functions and infrared fixed point in Landau gauge SU(N) Yang-Mills
%theory,''
Phys.\ Lett.\ B {\bf 611} (2005) 279
[arXiv:hep-th/0412330].

\bibitem{Llanes-Estrada:2004jz}
F.~J.~Llanes-Estrada, C.~S.~Fischer and R.~Alkofer,
%``Semiperturbative construction for the quark gluon vertex,''
arXiv:hep-ph/0407332.
%%CITATION = HEP-PH 0407332;%%


\end{thebibliography}
\end{document}